\documentclass[aps,prl,preprint,groupedaddress, showpacs]{revtex4-1}

\usepackage{graphics}
\usepackage{graphicx}
\usepackage{braket}
\usepackage{amsmath}
\usepackage{amssymb}

\begin{document}

\title{Ultrafast Reversal of a Fano Resonance in a Plasmon-Exciton System}


\author{Raman A.~Shah}

\author{Norbert F.~Scherer}
\affiliation{Department of Chemistry and The James Franck Institute,
  The University of Chicago, 929 East 57th Street, Chicago, Illinois
  60637, United States}

\author{Matthew Pelton}

\author{Stephen K.~Gray}
\email[]{gray@anl.gov}
\affiliation{Center for Nanoscale Materials, Argonne National
  Laboratory, 9700 South Cass Avenue, Argonne, Illinois 60439, United
  States}

\date{\today}

\begin{abstract}
When a two-level quantum dot and a plasmonic metal nanoantenna are
resonantly coupled by the electromagnetic near field, the system can
exhibit a Fano resonance, resulting in a transparency dip in the
optical spectrum of the coupled system. We calculate the nonlinear
response of such a system, for illumination both by continuous-wave
and ultrafast pulsed lasers, using both a cavity quantum electrodynamics and a
semiclassical coupled-oscillator model. For the experimentally
relevant case of meV thermal broadening of the quantum-dot transition,
we predict that femtosecond
 pulsed illumination can lead to a reversal
of the Fano resonance, with the induced transparency changing into a
superscattering spike in the spectrum. This ultrafast reversal is due
to a transient change in the phase relationship between the dipoles of
the plasmon and the quantum dot.  
It thus represents a new approach to dynamically control the
collective optical properties and coherence of coupled nanoparticle
systems.
\end{abstract}


\maketitle


A hybrid system of a semiconductor quantum dot (QD) in the near field
of a plasmonic metal nanostructure can exhibit qualitatively different
optical properties than its individual components \cite{matt_review,
  giannini_review, ming}. For sufficiently strong resonant coupling
between the QD exciton and the plasmon, the optical spectrum can
exhibit Fano interference \cite{sasha_prl, xiaohua_dit}: the QD creates 
a dramatic ``dipole-induced transparency,''
\cite{waks-vuckovic} suppressing absorption and scattering in spite of
its relatively meager oscillator strength.  A classical model can describe
this effect in the linear response limit.  Calculating the nonlinear
response of the hybrid system requires that the QD, at least, be modeled quantum
mechanically. Semiclassical (SC) models that treat the QD as a
two-level system have predicted novel nonlinear-optical effects, including a
``nonlinear Fano effect'' and optical bistability \cite{sasha_prl,
  artuso-bryant-2008, artuso-bryant-2010, artuso-bryant-2011,
  cheng_optics_letters}. Treating both plasmon and QD quantum
mechanically further refines the picture \cite{edo_cqed, sasha_prb,
  savasta_cqed, nordlander_quantum_plexcitonics,agio}, predicting in
particular a suppression of induced transparency, and of bistability,
due to additional dephasing not accounted for in the SC model
\cite{trugler-hohenester, savasta_cqed}.

These phenomena have all been predicted in the regime of very narrow
QD linewidths, on the order of $10\, \mu \text{eV} $, corresponding to
liquid-helium temperatures. Although this can be realistic when QDs
are coupled to practically lossless components such as
photonic-crystal cavities, absorptive heating may render such
temperatures infeasible for QDs coupled to plasmonic
nanostructures. We therefore consider here the more experimentally
achievable parameter regime of meV QD linewidths.  

In this regime, we predict a new nonlinear phenomenon for femtosecond
pulsed excitation; namely, for particular fluences the Fano resonance
reverses, resulting in a coherently enhanced cross section rather than
an induced transparency. This is a dynamical analog of the transition
from electromagnetically induced transparency to superscattering
\cite{verslegers}, arising from a transient change in the phase
relationship between the QD and the plasmon. It is thus fundamentally
different from the nonlinear Fano effect, a steady-state change in the
Fano lineshape arising from the dependence of the QD-plasmon coupling
strength on the incident field intensity \cite{sasha_prl, sasha_prb}.

Our treatment starts with a quantum-mechanical model of the
hybrid system.  We follow a previously developed cavity quantum electrodynamics (CQED)
formalism \cite{edo_cqed, sasha_prl, savasta_cqed}, extending the previous work to the
regime of broader QD linewidths and to consideration of the
transient optical response.  The underlying basis states are
$\Ket{qs}$, where $ q \in \set{0,1} $ indexes the QD energy levels and
$ s \in \set{0, 1, 2, \ldots} $ indexes plasmon energy levels.
Lowering and raising operator pairs for the QD and plasmon are
$(\hat{\sigma} , \hat{\sigma}^+)$ and $(\hat{b},\hat{b}^+)$,
respectively.  The dipole operators are then $ {\hat{\mu}_q} = d_q (
{\hat{\sigma}} + \hat{{\sigma}}^{+} ) $ and $ {\hat{\mu}_s} = d_s (
\hat{b} + \hat{b}^{+} ) $, where $ d_q $ and $ d_s $ are the
transition dipole moments of the QD and plasmon, respectively, and the
total dipole operator is $ \hat{\mu} = \hat{\mu_s} + \hat{\mu_q} $.
The evolution of the density operator $ \hat{\rho}(t) $ is governed by
\begin{equation}\label{master_eqn}
\frac{\mathrm{d} \hat{\rho}}{\mathrm{d} t} 
= -\frac{i}{\hbar} [ \hat{H}, \hat{\rho} ] +
L(\hat{\rho})~~~~,
\end{equation}
where $ \hat{H} $ is the system Hamiltonian,
and $ L(\hat{\rho}) $ is a Lindblad superoperator 
accounting for disspiation.
Specifically, 
$\hat{H} = \hat{H_q} + \hat{H_s} + \hat{H_i} + \hat{H_d}$,
where $ \hat{H_q} = \hbar \omega_q \hat{\sigma}^{+} \hat{\sigma}
$ is the QD Hamiltonian, $ \hat{H_s} = \hbar \omega_s
\hat{b}^{+} \hat{b} $ is the plasmon Hamiltonian, $
\hat{H_i} = - \hbar g ( \hat{\sigma}^{+} {b} + \hat{\sigma}
{b}^{+} ) $ describes plasmon-QD coupling, and $
\hat{H_d} = - E(t) \hat{\mu} $ describes driving by an incident
electric field $ E(t) $ \cite{suppinfo}. 
The Lindblad superoperator is \cite{edo_cqed} 
\begin{equation}\label{lindblad_q}
L_q(\hat{\rho}) = 
- \frac{\gamma_1}{2} 
(\hat{\sigma^{\dagger}} \hat{\sigma} \hat{\rho} 
+ \hat{\rho} \hat{\sigma^{\dagger}} \hat{\sigma} 
- 2 \hat{\sigma} \hat{\rho} \hat{\sigma^{\dagger}}) 
- \gamma_2 
(\hat{\sigma^{\dagger}} \hat{\sigma} \hat{\rho} 
+ \hat{\rho} \hat{\sigma^{\dagger}} \hat{\sigma} 
- 2 \hat{\sigma^{\dagger}}
   \hat{\sigma} \hat{\rho} \hat{\sigma^{\dagger}} \hat{\sigma}) \, .
\end{equation}
The first term describes spontaneous emission with rate $\gamma_1 = T_1^{-1} $,
and the second term describes dephasing with rate
$ \gamma_2 = T_2^{-1} $.
In order to solve Eq. (\ref{master_eqn}), we define 
a maximum number of plasmons, $ N_s $, above which the
density matrix elements are negligible. Solution then 
involves integrating $ O(N_s^2) $ coupled ordinary
differential equations, or, at steady state, $ O(N_s^2) $
coupled algebraic equations \cite{suppinfo}. Once a
solution is obtained, 
the total dipole is calculated according to $ \mu(t) =
\mathrm{Tr} \left( \hat{\rho}(t) \hat{\mu} \right)$.

A computationally simpler approach is a SC or Maxwell-Bloch model, in which the
plasmon dipole, $\mu_s(t)$, is treated classically and the QD is
treated with Bloch equations \cite{ziolkowski} 
or their generalization \cite{shameless}
for its reduced density
matrix, $\rho^{QD}(t)$:
\begin{eqnarray}\label{sc_model}
\ddot{\mu_s} + \gamma_s \dot{\mu_s} + \omega_s^2 \mu_s
= A_s \left[ E + J \mu_q \right] \, , ~~~~~~~~~~~~~~~~~~~~  \nonumber \\
\dot{\rho}_1
= \omega_q \rho_2 - \gamma_2 \rho_1,
~~~~~
\dot{\rho}_2
= - \omega_q \rho_1 - \frac{2 d_q}{\hbar}
\left[ E + J \mu_s \right] \rho_3 - \gamma_2 \rho_2 \, ,
~~~~ \\
\dot{\rho}_3
=  \frac{2 d_q}{\hbar}
\left[ E + J \mu_s \right] \rho_2 - \gamma_1^{\text{SC}} ( \rho_3 + 1 ) \, ,
~~~~~~~~~~~~~~~~
\nonumber
\end{eqnarray}
where $E$ = $E(t)$, $\rho_1 = 2 \, \mathrm{Re} \rho^{QD}_{01}(t), $ $\rho_2 = -2 \,
\mathrm{Im} \rho^{QD}_{01}(t), $ $\rho_3 = \rho^{QD}_{11}(t) -
\rho^{QD}_{00}(t), $ and $\mu_q =  d_q \rho_1$.  One can relate $ A_s
$ to the CQED parameters by solving for the steady state of $\mu_s$
with $J$ = 0 \cite{suppinfo}, giving $A_s = {4 \omega_s d_s^2}/{\hbar}
$. Comparing classical and quantum dipole interaction energies gives $
J = {\hbar g}/(d_q d_s) $ \cite{suppinfo}.
Eqs.~\ref{sc_model} can be numerically integrated
with or without the rotating wave approximation \cite{suppinfo}; the
results are essentially identical for the present problem.  
The SC
model involves the solution of a fixed, small number of ordinary
differential equations regardless of intensity, unlike the CQED model,
which becomes computationally costly for large $N_s.$

A flaw in the SC model is that an excited QD in the dark
cannot couple to the plasmon, but does so in CQED 
via the Purcell effect \cite{wang-vyas,edo_cqed,suppinfo}.
With
$ \omega_q = \omega_s $ and $ \gamma_s \gg
\gamma_1, \gamma_2 $, and $ g $, 
the QD's 
effective decay constant due to the Purcell effect is \cite{suppinfo}
$\gamma_1^{\text{eff}} \approx {4g^2}/{\gamma_s}$.  
We therefore term the SC model with $ \gamma_1^{\text{SC}} $ = $\gamma_1$
as the ``na\"ive'' SC model, and with $\gamma_1^{\text{SC}} = $ $\gamma_1^{\text{eff}}$ as
the ``corrected'' SC model.  

Having obtained $\mu(t)$ using either the CQED or SC model, the
absorption cross section is found according to
\cite{ novotny-hecht,suppinfo}:
\begin{equation}\label{cross_section}
\sigma_{abs}(\omega) =  \frac{k}{\epsilon_0} {\rm Im} [ \alpha (\omega) ] \, ,
\end{equation}
where $ k $ = $ \sqrt{\epsilon_{med}} \, \omega / c$, with $\epsilon_{med}$ being
the relative dielectric constant of the surounding medium, and 
\begin{equation}\label{alpha_def}
\alpha(\omega) = 
\frac{\int{\mathrm{e}^{-i \omega t} \mu(t)\,\mathrm{d}t}}
{\epsilon_{med}\int{\mathrm{e}^{-i \omega t} E(t)\,\mathrm{d}t}} \, ,
\end{equation}
where integration is over an optical cycle after steady
state is reached or over all time for pulsed excitation. 
While we discuss absorption cross sections here, 
scattering spectra exhibit nearly identical trends.

\begin{figure}
  \includegraphics[width=100mm]{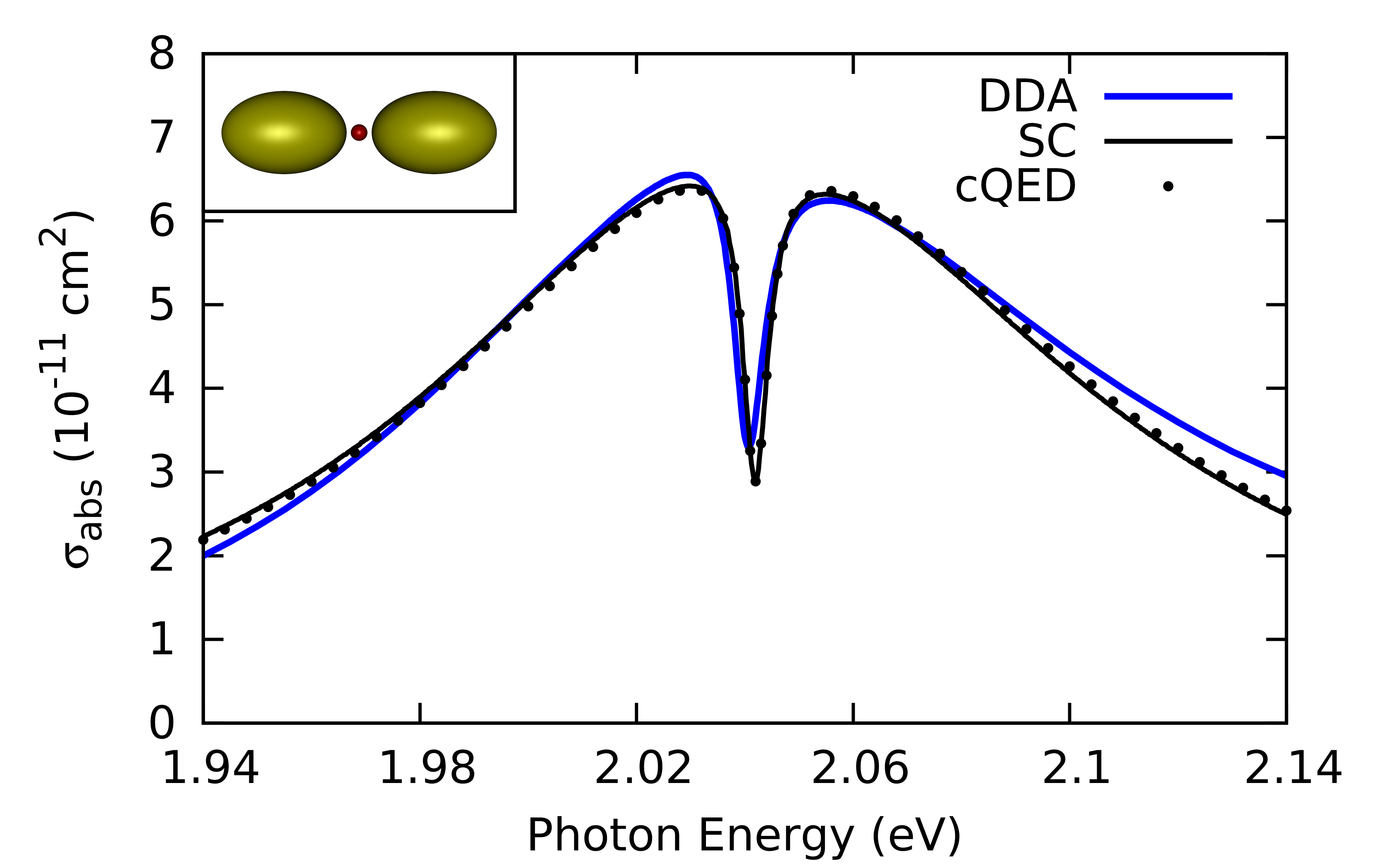}
  \caption{Linear absorption spectrum of an Au--CdSe--Au hybrid nanoparticle system
  (illustrated in the inset), calculated using the discrete dipole approximation (DDA),
  cavity-quantum-electrodynamics model (CQED), and semiclassical model (SC).}
  \label{fig1}
\end{figure}

The model parameters are obtained by fitting to spectra for a
realistic system calculated with the discrete dipole approximation
(DDA) \cite{draine-flatau,suppinfo}. As illustrated in Fig.~\ref{fig1}, two Au prolate
spheroids with semi-major and semi-minor axes of 15 nm and 10 nm,
respectively, are arranged coaxially with a gap of 6 nm. A 4 nm
diameter CdSe QD is placed in the center of the gap. The system is
embedded in a medium with dielectric constant $ \epsilon_{med} $ =
2.25, typical of a polymer or glass.  The QD dielectric constant is
taken to be a Lorentzian function  with center frequency chosen to
match the plasmon frequency of the metal nanostructure, and with
linewidth corresponding to temperatures of 
50 -- 100 K \cite{xiaohua_dit, besombes-kheng,suppinfo}.  The
fitting gives $ \hbar \omega_s = 2.042 $ eV, $ d_s = 2990 $ D, $ \hbar
\gamma_s = 150 $ meV; $ \hbar \omega_q = 2.042 $ eV, $ d_q = 13.9 $ D,
$ \hbar \gamma_2 = 1.27 $ meV; and $ \hbar g = 10.8 $ meV \cite{suppinfo}.  As seen in
Fig.~\ref{fig1}, the CQED and SC results are in excellent agreement
with each other and in good agreement with the DDA spectrum.  
The QD spontaneous emission rate is
calculated according to $\gamma_1 =$ $ \omega_q^3
\sqrt{\epsilon_{med}} d_q^2/(3 \pi \epsilon_0 \hbar c^3)$
\cite{nienhuis-alkemade}, giving $ \hbar \gamma_1 $ = 268 neV or $ T_1
= \gamma_1^{-1} = 2.46 $ ns; this, in turn, gives $ \hbar
\gamma_1^{\text{eff}} = 3.02 $ meV, corresponding to $
T_1^{\text{eff}} = 218 $ fs.

\begin{figure}
  \includegraphics{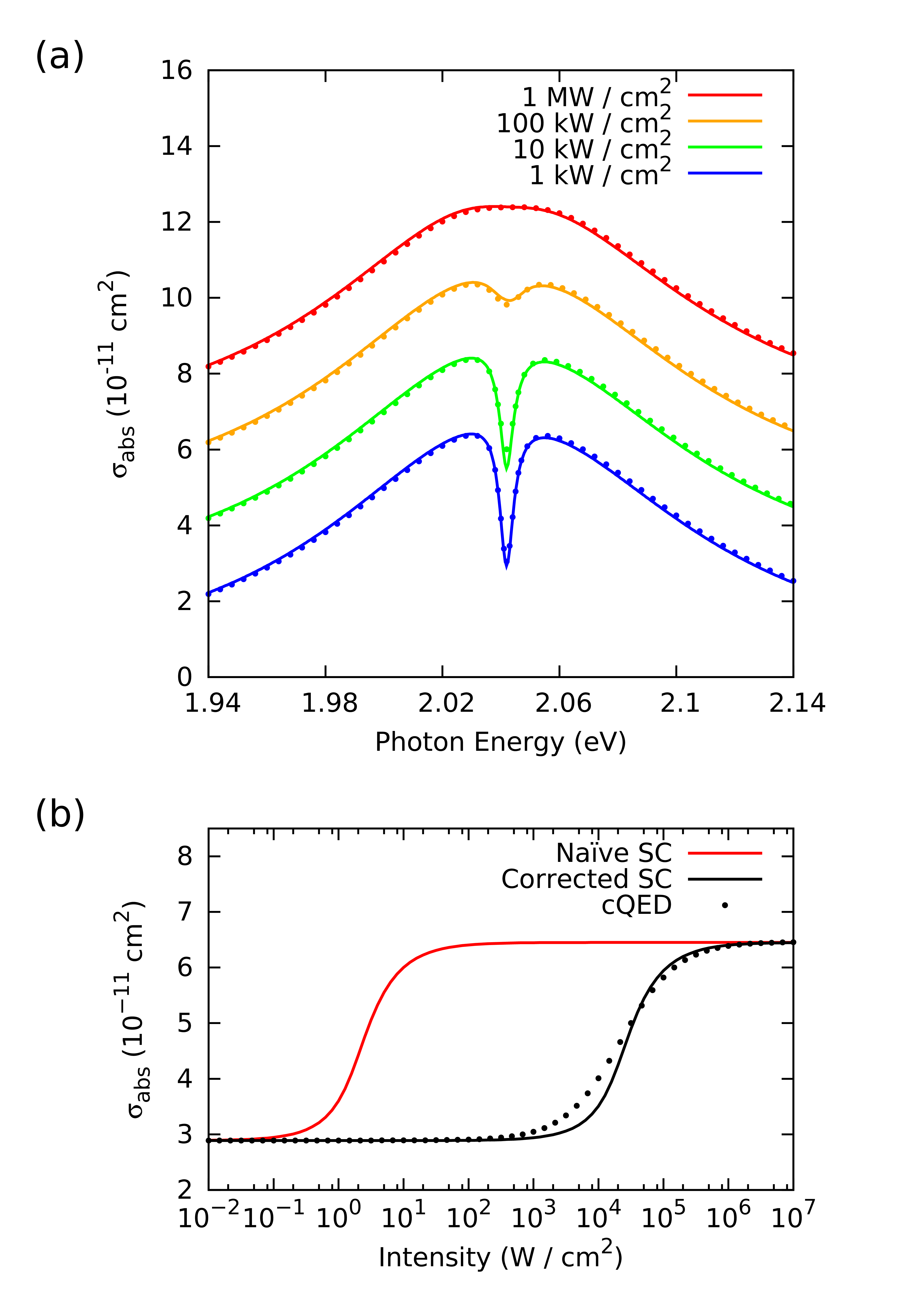}
  \caption{(a) Steady-state absorption spectra calculated using the CQED (dots) and corrected SC (solid)
  model for various incident intensities. 
    Successive spectra are displaced vertically by $ 2 \times 10^{-11}\,
    \text{cm}^2 $. (b) Intensity dependence of steady-state absorption cross-sections
    calculated using the
    CQED, na\"ive SC and corrected SC steady-state models, at a photon energy of
    2.042 eV.} 
  \label{fig2}
\end{figure}

We begin by considering steady-state spectra. Fig.~\ref{fig2} shows that the Fano
resonance dip disappears as the incident intensity is increased, 
due to saturation of the QD transition.  The corrected SC
and CQED models are in excellent quantitative agreement for high and
low applied fields.  The na\"ive SC model is gravely in error,
but can be brought into agreement with the corrected SC model by
multiplying the incident intensity by the Purcell factor \cite{edo_cqed,agio}.  This validates
the use of the SC model, which is particularly important for
simulations at the high intensities for which CQED calculations
are computationally prohibitive.  The results for the current system
contrast with predictions for systems with narrow QD linewidths; for these systems,
the SC formulation gives a deeper transparency even in the linear
regime, due to quantum-optical dephasing that is ignored by the
SC model \cite{edo_cqed}. In our system, thermal dephasing
dominates over this vacuum-field-induced dephasing. As the field is
increased, however, the quantum-optical dephasing increases and
eventually becomes comparable to the thermal linewidth, resulting in
a small disagreement between our CQED and SC predictions at moderate
fields.

\begin{figure}
  \includegraphics{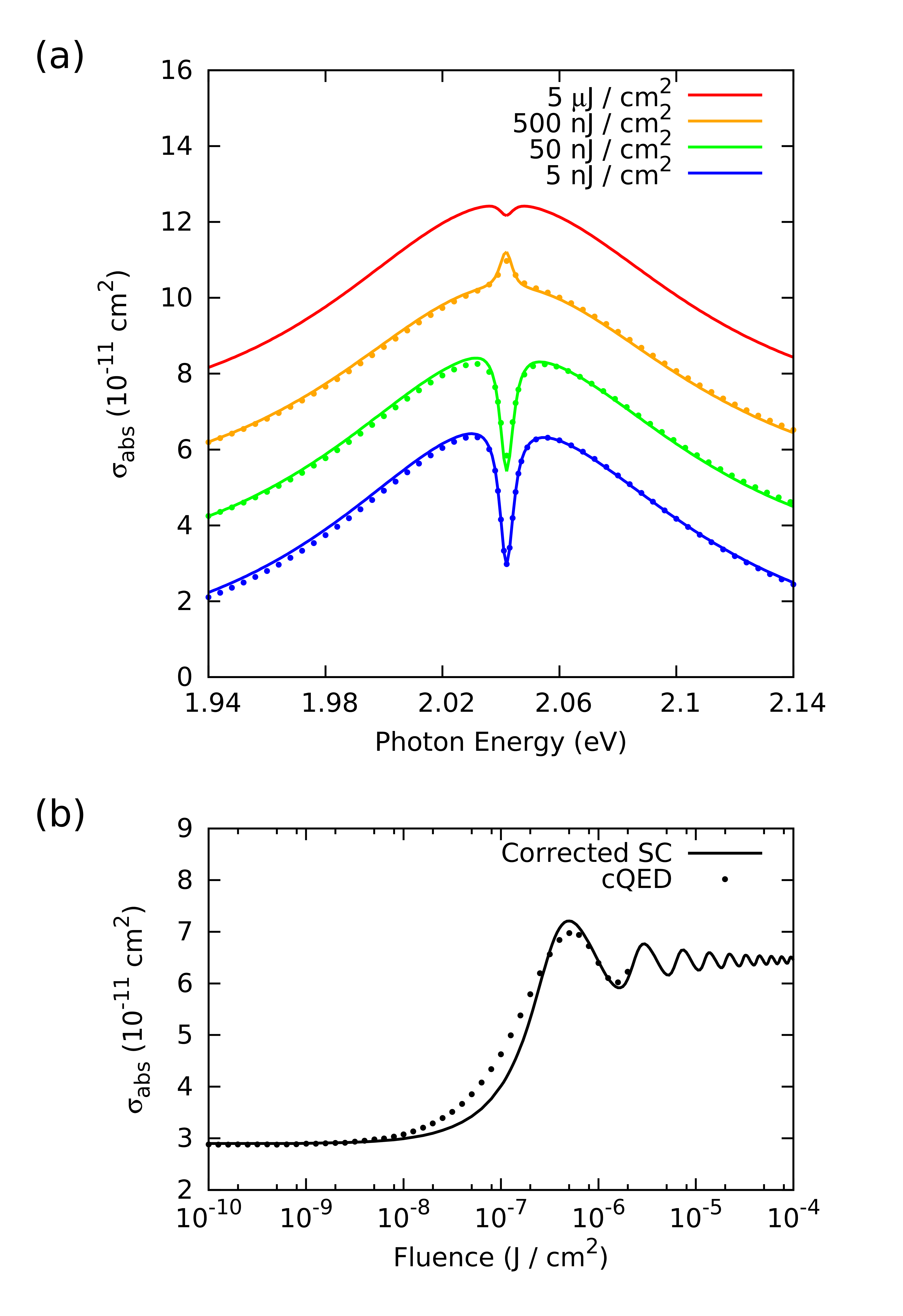}
  \caption{(a) Absorption spectra calculated using the CQED (dots) and corrected SC (solid) 
   models, for ultrafast pulsed excitation. 
     Successive spectra, corresponding to varying fluences,
     are displaced vertically by $ 2 \times 10^{-11}\,
    \text{cm}^2 $. (b) Fluence dependence of the absorption cross-section
    for ultrafast pulsed excitation, calculated using the CQED and SC models, at a
    photon energy 2.042 eV.} 
  \label{fig3}
\end{figure}

We next consider the system response to a Gaussian
pulse, $E(t)$, that has a 20 fs
full width at half the maximum intensity and a center frequency, $
\hbar \omega = 2.042 $ eV, resonant with the plasmon and exciton.
Transient spectra obtained by Fourier transformation of the time-domain
response are shown in Figure \ref{fig3}.
At low intensities, the transient spectrum
is identical to the linear steady-state spectrum. 
Strikingly, however, at
certain higher fluences the resonant dip reverses to form a narrow spike.
The corrected SC model remains in excellent
agreement with the CQED model. As the pulse fluence increases beyond
the range that is readily computable using the CQED model, 
the SC model predicts recurrences of the
ultrafast reversal suggestive of Rabi oscillations.

\begin{figure}
  \includegraphics{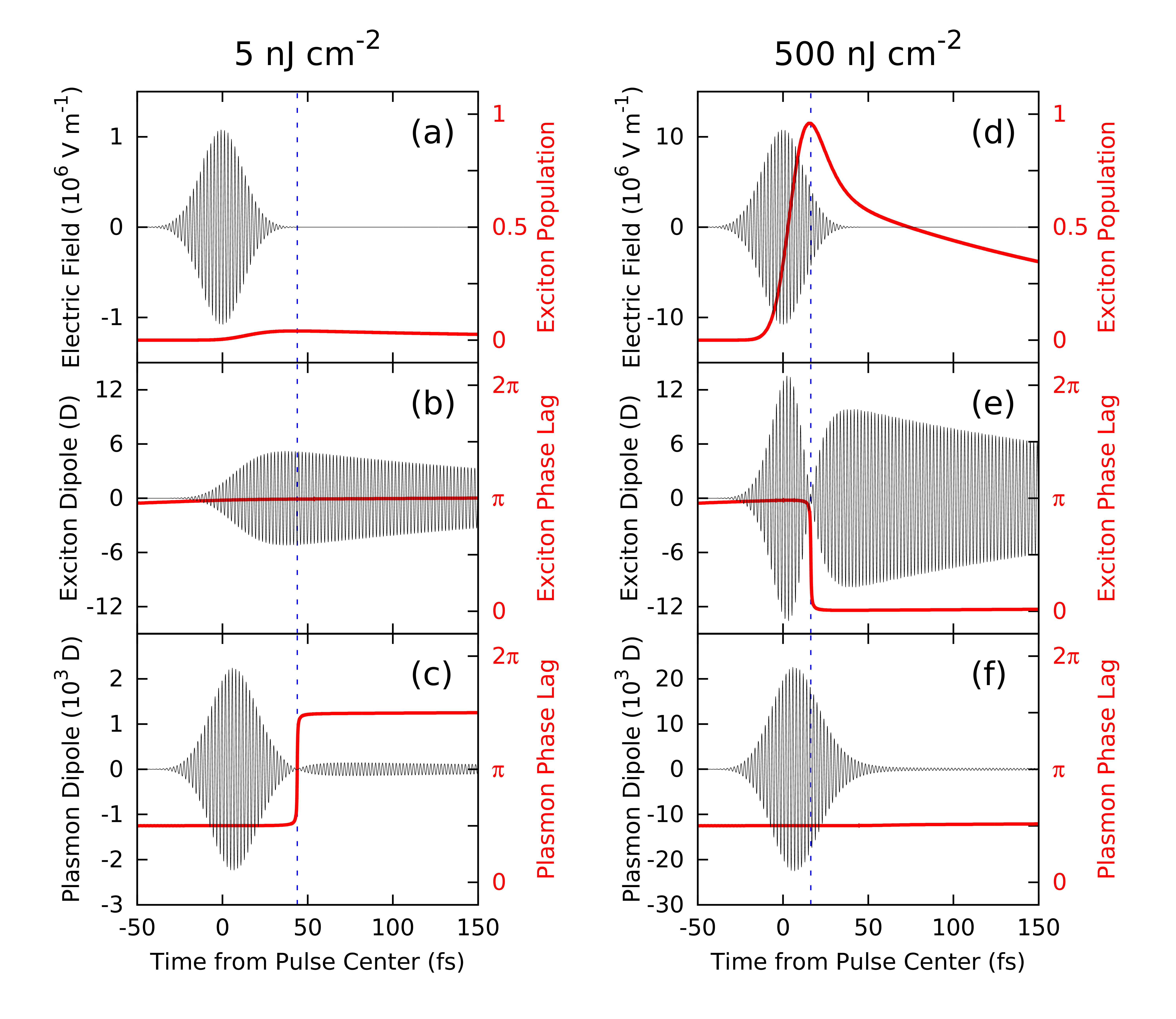}
  \caption{Time evolution of the QD-plasmon system 
     under pulsed excitation, calculated using the CQED model
     for two fluences.
     (a,d) Pulse's electric field and
    QD population. (b,e) QD's dipole and phase lag relative to the pulse. 
    (c,f) Plasmon's dipole  and
    phase lag. 
    Dashed lines indicate times at which a phase jump
    occurs.}
  \label{fig4}
\end{figure}

The ultrafast reversals of the Fano resonance arise from the transient
phase relationships of the plasmon and QD dipoles with respect to the
incident light, as illustrated in Fig.~\ref{fig4}. At the beginning of
the incident pulse, as at steady state, the plasmon lags the driving laser by the
${\pi}/{2}$ phase difference expected for a resonant oscillator \cite{agio}.  The QD is driven
primarily by the plasmon and thus lags the laser by an additional $
{\pi}/{2}$ phase difference. At low fluence, this phase relationship continues
until the pulse is complete and the short-lived plasmon has
decayed. Then, the longer-lived QD, still oscillating with a $ \pi $
phase lag relative to the laser, drives the plasmon; the plasmon thus
acquires a $ \pi + ({\pi}/{2}) = ({3 \pi}/{2})$ phase lag, partially
canceling its earlier oscillations in the spectral domain and
producing the observed linear Fano dip. By constrast, at higher
fluences, the QD population exhibits Rabi oscillations, reaching unity
and then coherently being driven back down. This reverses the sign of
the QD dipole \cite{boyd}, so that the lag of the QD phase relative to
the laser is now zero.  The phase frustration that previously led to
transparency is replaced by a constructive interference that leads to
induced absorption or superscattering.

Ultrafast reversal is thus due to a change in the phase of the
coherent interaction between the QD and the plasmonic metal
nanostructure. A similar change has been demonstrated in plasmonic or
metamaterial systems that exhibit Fano resonances due to the coupling
between bright modes and dark modes \cite{luk'yanchuk}. In this case,
the sign of the interference can be controlled through careful
selection of the linewidths and coupling strengths \cite{tassin}, or
by adding a retardation-based phase delay \cite{taubert}. 
Similarly, a change from steady-state transparency to enhanced
absorption has been predicted in coupled QD-plasmon systems by 
changing the size of the metal nanoparticle and the detuning \cite{agio}.
In these
systems, reversal from destructive interference to constructive
interference can be controlled only statically, by changing the
structure of the system. In our QD-plasmon system, by contrast, the
reversal can be controlled dynamically, by changing the fluence of the
incident ultrafast pulses.

Not all QD-plasmon systems that exhibit Fano resonances will exhibit
ultrafast reversal. The reversal requires that the QD dipole
oscillates significantly longer than the plasmon's intrinsic lifetime.
In other words, we require $\gamma_1^{\text{eff}} \ll \gamma_s $,
which implies $ g \ll \gamma_s/2 $.  However, $ g $ must also be large
enough to give a Fano resonance, which requires $ g \ge \sqrt{\gamma_s
  \gamma_2}/4$ \cite{xiaohua_dit,suppinfo}.  These constraints are
more easily satisfied for $ \gamma_2 \ll \gamma_s $. The large $
\gamma_s $ afforded by a lossy plasmonic component points to the
intrinsically plasmonic nature of ultrafast reversal: such constraints
are unlikely to be satisfied by a high-finesse photonic resonator.

Sufficiently narrow QD linewidths can be obtained by cooling to
liquid-nitrogen temperatures, even if we account for
absorption-induced heating of the plasmonic nanostructure. Achieving
the required coupling strengths is a greater experimental challenge,
but should be feasible using chemically synthesized components and
directed assembly, such as DNA-based assembly of colloidal QDs and
metal nanoparticles \cite{cohen-hoshen}. The availability of low-cost
assembly methods is an advantage for these systems as compared to
traditional CQED systems, which require complex and expensive top-down
fabrication \cite{englund, fushman}.

Our treatment of the optical response of coupled QD-metal
nanostructure systems has employed a larger thermal dephasing rate for
the QD than has generally been considered in previous treatments.
Although this means that certain phenomena requiring a high degree
of coherence are suppressed,
significant quantum-optical effects remain.  First, the saturation of
the Fano resonance is the principal optical nonlinearity at steady
state, and the intensity at which this saturation occurs is due to a
balance between two inextricable aspects of QD-plasmon coupling:
plasmonic field enhancement lowers the incident fields required for
saturation, while the Purcell effect increases the required
fields. Second, we
predict that the Fano resonance can undergo a reversal, changing from
a transparency dip into a superscattering spike, when excited by
femtosecond laser pulses with appropriate fluence.  This ultrafast
reversal represents a new means to coherently control optical
interactions among nanostructures.

\begin{acknowledgments}
We thank Dr.~Lina Cao for helpful conversations. This research was
funded by the National Science Foundation (CHE-1059057).  R.~A.~S.~was
supported by a National Science Foundation Graduate Research
Fellowship. Use of the Center for Nanoscale Materials was supported by
the U.~S.~Department of Energy, Office of Science, Office of Basic
Energy Sciences User Facility, under Contract No.~DE-AC02-06CH11357.
\end{acknowledgments}

\bibliography{dit}

\end{document}